\documentclass[aps,showpacs,preprintnumbers,amsmath, amssymb]{revtex4}

\usepackage{float}
\usepackage{graphics,epsfig}
\usepackage{graphicx}
\usepackage{dcolumn}
\usepackage{bm}

\begin{document}
\baselineskip=0.8 cm

\title{{\bf Upper bound on the radius of the innermost photonsphere in the regular compact star spacetime}}
\author{Guohua Liu$^{1}$\footnote{liuguohua1234@163.com}}
\author{Yan Peng$^{2}$\footnote{yanpengphy@163.com}}
\affiliation{\\$^{1}$ School of Physics and Physical Engineering, Qufu Normal University, Qufu, Shandong, 273165, China}
\affiliation{\\$^{2}$ School of Mathematical Sciences, Qufu Normal University, Qufu, Shandong 273165, China}

\vspace*{0.2cm}
\begin{abstract}
\baselineskip=0.6 cm
\begin{center}
{\bf Abstract}
\end{center}

We study properties of the innermost photonsphere in the regular compact star background.
We take the traceless energy-momentum tensor and dominant
energy conditions. In the regular compact star background,
we analytically obtain an upper bound on the radius of the innermost photonsphere
as $r_{\gamma}^{in}\leqslant \frac{12}{5}M$, where $r_{\gamma}^{in}$
is the radius of the innermost photonsphere and $M$ is the total ADM
mass of the asymptotically flat compact star spacetime.

\end{abstract}

\maketitle
\newpage
\vspace*{0.2cm}

\section{Introduction}

A characteristic property of highly curved spacetimes is the existence of
null geodesics \cite{c1,c2}. The circular null geodesics are usually called photonspheres,
which correspond to the trajectory of massless particles traveling around central objects.
The photonsphere can provide information on the curved spacetime and
thus, have attracted lots of attention from mathematical and physical aspects
 \cite{ad1,ad2,ad3,ad4,ad5,c3,c4,c5,z1,z2,z3}.
It is closely related to novel properties,
such as the strong gravitational lensing phenomenon \cite{c6},
distributions of exterior matter fields \cite{s1,s2,s3,s4,s5,s6},
circular period around central objects \cite{s7,s8,s9},
nonlinear instabilities of spacetimes \cite{us1,us2,us3,us4,us5,us6,us7}
and characteristic resonances of black holes \cite{r1,r2,r3,r4,r5,r6,r7,r8}.

An interesting question is how close can the photonspheres be to the central compact objects.
For a generic black hole spacetime, Hod analytically derived an upper bound
on the radius of the outermost photonsphere with non-positive
trace of the energy-momentum tensor $T\leqslant0$. The bound is expressed as
$r_{\gamma}^{out}\leqslant 3M$, where $r_{\gamma}^{out}$
is the radius of the outermost photonsphere and $M$
is the total ADM mass of the black hole spacetime \cite{b3}.
In the background of horizonless ultra-compact stars, it was found that
the radius of the photonsphere is bounded from above by $r_{\gamma}^{out}\leqslant 4M$ for the dominant energy
condition and a stronger upper bound $r_{\gamma}^{out}\leqslant \frac{24}{7}M$
was obtained for negative isotropic trace $T<0$, where $r_{\gamma}^{out}$
is the radii of outermost photonspheres of compact stars and $M$
is the total ADM mass of the compact star spacetime \cite{b3}.
In the traceless energy-momentum tensor case $T=0$, it was proved that
the innermost photonsphere cannot be arbitrarily close to the central black hole
with a lower bound $r_{\gamma}^{in}\geqslant \frac{6}{5}r_{H}$,
where $r_{\gamma}^{in}$ is the radius of the innermost photonsphere and $r_{H}$
is the horizon of the black hole \cite{Hod}.
As a further step, it is also interesting to study properties of
innermost photonspheres of horizonless compact stars.

In this work, we firstly introduce the gravity system of horizonless compact stars.
Then we investigate on properties of innermost
photonspheres surrounding compact stars.
With analytical methods, we obtain an upper bound
on the radii of innermost photonspheres of compact stars.
We will summarize our main results in the last section.

\section{The gravity system in the compact star spacetime}

We are interested in spherically symmetric horizonless compact stars with
photonspheres. The curved spacetime is characterized by \cite{c2,s1,s3}
\begin{eqnarray}\label{AdSBH}
ds^{2}&=&-\mu(r)e^{-2\delta(r)}dt^{2}+\mu^{-1}dr^{2}+r^{2}(d\theta^2+sin^{2}\theta d\phi^{2}),
\end{eqnarray}
where $\mu(r)$ and $\delta(r)$
are radially dependent metric functions.

At the original center, the regular spacetime satisfies
\cite{b1,b2}
\begin{eqnarray}\label{AdSBH}
\mu(r\rightarrow 0)=1+O(r^2)~~~~~~and~~~~~~\delta(0)<\infty.
\end{eqnarray}

At the infinity, the asymptotic flatness requires
\begin{eqnarray}\label{AdSBH}
\mu(r\rightarrow \infty)=1~~~~~~and~~~~~~\delta(r\rightarrow \infty)=0.
\end{eqnarray}

We define $\rho$, $p$ and $p_{\tau}$ as the energy density
$\rho=-T_{t}^{t}$, radial pressure $p=T_{r}^{r}$ and
tangential pressure $p_{\tau}=T_{\theta}^{\theta}=T_{\phi}^{\phi}$ respectively.
The equations of motion for metric functions are
\begin{eqnarray}\label{BHg}
\mu'=-8\pi r \rho+\frac{1-\mu}{r},
\end{eqnarray}
\begin{eqnarray}\label{BHg}
\delta'=\frac{-4\pi r (\rho+p)}{\mu}.
\end{eqnarray}

We take the dominant energy condition
\begin{eqnarray}\label{BHg}
\rho\geqslant |p|,~|p_{\tau}|\geqslant 0.
\end{eqnarray}

We also assume the traceless energy-momentum tensor condition
\begin{eqnarray}\label{BHg}
T=-\rho+p+2p_{\tau}=0.
\end{eqnarray}

The gravitational mass contained in a sphere with radius r
is given by
\begin{eqnarray}\label{AdSBH}
m(r)=\int_{0}^{r}4\pi r'^{2}\rho(r')dr'.
\end{eqnarray}

Without generality, the metric function $\mu(r)$
can be putted in the form \cite{b2}
\begin{eqnarray}\label{BHg}
\mu(r)=1-\frac{2m(r)}{r}.
\end{eqnarray}

According to (8), the finiteness of the gravitational mass requires \cite{b1}
\begin{eqnarray}\label{AdSBH}
r^{3}\rho(r)\rightarrow 0~~~~~~as~~~~~~r\rightarrow\infty.
\end{eqnarray}

\section{Upper bounds on radii of the innermost photonspheres}

In this part, we will obtain an upper bound on radii of
innermost photonspheres of horizonless compact stars.
We get the characteristic equation of photonspheres
following analysis in \cite{c2,s3,us2}.
The independence of the metric (1) on both t and $\phi$
leads to the conserved energy E and conserved angular momentum L.
The photonspheres of the spherically symmetric star is described by
\begin{eqnarray}\label{BHg}
V_{r}=E^{2}~~~~~~and~~~~~~V_{r}'=0,
\end{eqnarray}
where $V_{r}$ is the effective radial potential
\begin{eqnarray}\label{BHg}
V_{r}=(1-e^{2\chi})E^{2}+g\frac{L^2}{r^2}.
\end{eqnarray}

Substituting (4) and (5) into
(11) and (12), we get the photonsphere equation \cite{b1}
\begin{eqnarray}\label{BHg}
N(r)=3\mu(r)-1-8\pi r^2p=0.
\end{eqnarray}

The roots of characteristic equation (13) satisfying
\begin{eqnarray}\label{BHg}
N(r_{\gamma})=3\mu(r_{\gamma})-1-8\pi (r_{\gamma})^2p=0
\end{eqnarray}
are the discrete radii of the photonspheres.

According to relations (3), (6) and (10), the radial
function $N(r)$ has the boundary behavior
\begin{eqnarray}\label{BHg}
N(r=0)=2~~~~~~and~~~~~~N(r\rightarrow\infty)\rightarrow2.
\end{eqnarray}

We are interested in compact stars with photonspheres.
We define $r_{\gamma}^{in}$ as the innermost photonsphere of the regular ultra-compact star.
From equation (15), one deduces that the innermost photonsphere satisfies the relation \cite{b1,b2,us5}
\begin{eqnarray}\label{BHg}
N'(r=r_{\gamma}^{in})\leqslant0.
\end{eqnarray}

The conservation equation $T^{\mu}_{\nu;\mu}=0$ has only one nontrivial component
\begin{eqnarray}\label{BHg}
T^{\mu}_{r;\mu}=0.
\end{eqnarray}

Substituting equations (4) and (5) into (17), we get the radial pressure equation
\begin{eqnarray}\label{BHg}
p'(r)=\frac{1}{2r\mu}[(3\mu-1-8\pi r^2p)(\rho+p)-2\mu \rho -6\mu p+4 \mu p_{\tau}].
\end{eqnarray}

The relation (18) can be transformed into
\begin{eqnarray}\label{BHg}
(r^2p)'=\frac{r}{2g}[N(\rho+p)+2\mu(-\rho+p+2p_{\tau})]
\end{eqnarray}
with $N=3\mu-1-8\pi r^2p$.

From relations (4), (7), (14) and (19), we get the equation \cite{b1,b2,us5}
\begin{eqnarray}\label{BHg}
N'(r=r_{\gamma}^{in})=\frac{2}{r_{\gamma}^{in}}[1-8\pi (r_{\gamma}^{in})^{2}(\rho+p_{T})].
\end{eqnarray}
Putting (20) into (16), we obtain the inequality
\begin{eqnarray}\label{BHg}
1-8\pi (r_{\gamma}^{in})^{2}(\rho+p_{T})=1-12\pi (r_{\gamma}^{in})^{2}\rho+4\pi (r_{\gamma}^{in})^{2}p\leqslant0
\end{eqnarray}
at the innermost photonsphere of the compact star.

From (6) and (21), we deduce the relation \cite{b2,b3}
\begin{eqnarray}\label{BHg}
1+16\pi (r_{\gamma}^{in})^{2}p\leqslant0.
\end{eqnarray}

Considering relations (14) and (22), we arrive at the inequality
\begin{eqnarray}\label{BHg}
6\mu(r_{\gamma}^{in})-1\leqslant0.
\end{eqnarray}

With the expression (9), we can transform (23) into
\begin{eqnarray}\label{BHg}
\frac{m(r_{\gamma}^{in})}{r_{\gamma}^{in}}\geqslant \frac{5}{12}.
\end{eqnarray}

We further obtain an upper bound for the radius of the innermost photonsphere as \cite{b2}
\begin{eqnarray}\label{BHg}
r_{\gamma}^{in}\leqslant \frac{12}{5}m(r_{\gamma}^{in})\leqslant \frac{12}{5}M.
\end{eqnarray}

\section{Conclusions}

We studied innermost photonspheres of horizonless compact stars
in the asymptotically flat background.
We assumed the traceless energy-momentum tensor and dominant
energy conditions for matter fields.
With analytical methods, we obtained an upper bound on the radii of innermost photonspheres
as $r_{\gamma}^{in}\leqslant \frac{12}{5}M$, where $r_{\gamma}^{in}$
is the radii of the innermost photonspheres and $M$ is the total ADM
mass of the asymptotically flat compact star spacetime.
It means that the innermost photonspheres cannot be too far from the
original center of the horizonless compact stars.

\begin{acknowledgments}

This work was supported by the Shandong Provincial Natural Science Foundation of China under Grant
No. ZR2022MA074. This work was also supported by a grant from Qufu Normal University
of China under Grant No. xkjjc201906, the Youth Innovations and Talents Project of Shandong
Provincial Colleges and Universities (Grant no. 201909118), Taishan Scholar Project of Shandong Province (Grant No.tsqn202103062)
and the Higher Educational Youth Innovation Science
and Technology Program Shandong Province (Grant No. 2020KJJ004).

\end{acknowledgments}

\end{document}